\documentclass[11pt,a4paper]{scrartcl}
\usepackage[latin1]{inputenc}
\usepackage[T1]{fontenc}
\usepackage{helvet}
\usepackage{geometry}
\usepackage[hidelinks]{hyperref}
\usepackage{graphicx}
\usepackage{todonotes}
\usepackage{subcaption}

\normalsize 
\begin{document}

\renewcommand\familydefault{\sfdefault}
\renewcommand\sfdefault{phv}
\normalfont
\pagenumbering{arabic}

\title{FAIR and Open Computer Science\\ Research Software}
\author{Wilhelm Hasselbring, Leslie Carr, Simon Hettrick,\\Heather Packer, Thanassis Tiropanis}

\maketitle

\begin{center}
\textbf{Abstract}
\end{center}
In computational science and in computer science, research software is a central asset for research. 
Computational science is the application of computer science and software engineering principles to solving scientific problems, whereas computer science is the study of computer hardware and software design.

The Open Science agenda holds that science advances faster when we can build on existing results. Therefore, research software has to be reusable for advancing science.
Thus, we need proper research software engineering for obtaining reusable and sustainable research software. This way, software engineering methods may improve research in other disciplines. However, research in software engineering and computer science itself will also benefit from reuse when research software is involved. 

For good scientific practice, the resulting research software should be open and adhere to the FAIR principles (findable, accessible, interoperable and repeatable) to allow repeatability, reproducibility, and reuse. Compared to research data, research software should be both archived for reproducibility and actively maintained for reusability. The FAIR data principles do not require openness, but research software should be open source software. Established open source software licenses provide sufficient licensing options, such that it should be the rare exception to keep research software closed. 

We review and analyze the current state in this area in order to give recommendations for making computer science research software FAIR and open. We observe that research software publishing practices in computer science and in computational science show significant differences.

\newpage

\section*{Key Insights}

\begin{itemize}
	\item For good scientific practice in computer science research, evaluating and publishing research software is gaining attention.
	\item For reproducibility and for reusability of research software, specific solution approaches are required: archives such as Zenodo serve for reproducibility, while development platforms such as GitHub serves for use, reuse and active involvement.
	\item Research software publishing practices in computer science and in computational science show significant differences: computational science emphasizes reproducibility, while computer science emphasizes reuse.
\end{itemize}

\section*{Research Software}

Research software is software that is employed in the scientific discovery process or a research object itself. 
Computational science (also scientific computing) involves the development of research software for model simulations and data analytics to understand natural systems answering questions that neither theory nor experiment alone are equipped to answer. Computational science is a multidisciplinary field lying at the intersection of mathematics and statistics, computer science, and core disciplines of science and research.

Despite the increasing importance of research software to the scientific discovery process, well-established software engineering practices are rarely adopted in computational science~\cite{CiSE2018}, but the computational science community has started to appreciate that software engineering is central to any effort to increase its research software productivity~\cite{Ruede2018}.
Computer science, in particular software engineering, may help with reproducibility and reuse to advance computational science. 

Publishing research software as open source is an established practice in science; a popular open source repository is GitHub~\cite{Borges2016}. 
Researchers also disseminate the code and data for their experiments as virtual machines on repositories such as DockerHub.\footnote{\url{https://hub.docker.com/}} Open source software practices have enabled the more general open science agenda.
Open Science principles affect the research life-cycle, in the way science is performed, its results -- including software -- published, assessed, discovered, and monitored, as will be discussed in the following section. 

\section*{Open Science}

Replicability and reproducibility are the ultimate standards by which scientific claims are judged~\cite{Peng2011}. 
Reproducibility and reuse of research can be improved by increasing transparency of the research process and products via an open science culture~\cite{Nosek2015}.

Of the variations of open science~\cite{Fecher2014}, in this paper, we consider the \textit{pragmatic} and \textit{infrastructure} views.
The \textit{pragmatic} view regards open science as a method to make research and knowledge dissemination more efficient. It thereby considers science and research as a process that can be optimized by, for instance, modularizing the process of knowledge creation, opening the scientific value chain, including external knowledge and allowing collaboration through online tools such as Github, etc. 
The \textit{infrastructure} view is concerned with the technical infrastructure that enables emerging research practices on the Internet, for the most part software tools and applications, as well as (high-performance) computing systems. 
An example of such infrastructures is the envisioned European Open Science Cloud EOSC~\cite{EOSC2016}.

\section*{Research Software Engineering for Sustainable Research Software}

These above-mentioned views of open science require software engineering to enable sustainable research software. We need proper research software \textit{engineering} for obtaining sustainable research software. Research Software Engineers (RSE) combine an intimate understanding of research with expertise in software engineering.\footnote{\url{https://rse.ac.uk/who/}} 
It is a relatively new role in academia, but a highly popular one that has shown significant growth in countries around the world.\footnote{\url{https://www.software.ac.uk/what-do-we-know-about-rses-results-our-international-surveys}} The RSE community has displayed a particular interest in adopting and promoting open science, and as such are perfectly placed to help researchers adopt FAIR and open software practices.

Many researchers and research software engineers spend significant time creating and contributing to software, a resource which is currently under-represented in the scholarly record. This creates some problems:
\begin{enumerate}
	\item Trust in research relies on the peer review system but without ready access to the software used to perform a given experiment or analysis, it is difficult for readers to check a paper's validity. 
	\item Lack of access to the software underlying research makes it significantly more difficult to build new research results on top of existing ones: it is difficult to `stand on the shoulders of giants'. 
	\item Promotion and hiring in academic research are highly dependent on building a portfolio of well-cited papers, and researchers whose main work is software development often have fewer research papers published.
\end{enumerate}
One of the biggest obstacles to making research software sustainable is ensuring appropriate credit and recognition for researchers who develop and maintain such software. 
The goal is to address the challenges that researchers face regarding \textit{sustainable} research software.
It is also essential to sustain software by sustaining its communities (researchers, developers, maintainers, managers, active users).

\section*{Approaches to Software Publishing}

The scientific paper is central to the research communication process. Guidelines for authors define what aspects of the research process should be made available to the community to evaluate, critique, reuse, and extend. Scientists recognize the value of transparency, openness, and reproducibility. However, it remains unclear, how this may be achieved with software.

Various journals allow one to add supplementary material to an article, which may include software. Such additional material is usually just put into zip archives, whose content is neither reviewed nor further explained with metadata information or other documentation. Thus, this may fail to promote reuse.

More publishing-oriented practices were also explored. For instance, Elsevier conducted the ``Executable Paper Grand Challenge''  to enhance how scientific information is used and communicated, addressing both computational science and computer science~\cite{GABRIEL2011}. Several projects presented their concepts of ``Executable Papers'' which were published in the corresponding conference proceedings. The example paper~\cite{MAZANEK2011} from this competition uses literate programming 
to present a Curry program within the paper. Thus, it contains the complete concise source code of their software application, which is directly executable, together with sufficient documentation to be understandable.

Research software may also be published in software journals such as JOSS\footnote{\url{https://joss.theoj.org/}}, 
JORS\footnote{\url{https://openresearchsoftware.metajnl.com/}}
or Software Impacts\footnote{\url{https://www.journals.elsevier.com/software-impacts}} 
but this is rarely adopted in computer science. 

Some research communities are also building online platforms for sharing research software services. The SoBigData Lab,\footnote{\url{http://sobigdata.eu}} for instance, provides a cloud service for data analytics, with a focus on social mining research. 
OceanTEA provides an online service for analyzing ocean observation data~\cite{CI2016}.
The integrated toolchain LabPal for running, processing, and including the results of computer experiments in scientific publications is presented in~\cite{Halle2018}. The tool Qresp for curating, discovering and exploring reproducible scientific papers is presented in~\cite{Govoni2019}.
Generic services such as BinderHub\footnote{\url{https://binderhub.readthedocs.io/}} support online execution of reproducible code. 

\section*{Artifact Evaluation as a Review Mechanism}

The code quality of research software often is a hindrance for reuse. For example, there are typically no tests, documentation is often lacking, and the code does not usually adhere to any coding standards. This is not necessarily caused by the scientist's bad work, but rather it is the natural result of what scientists are judged on, namely the scientific quality of the papers they put out, as opposed to the quality of software that enables such papers. 

To address these issues, several ACM conferences initiated artifact evaluation processes, in which supplementary material is part of the review process~\cite{Krishnamurthi2015}. 
The ACM distinguishes between repeatability (same team, same experimental setup), replicability (different team, same experimental setup), reproducibility (different team, different experimental setup), or reusability (artifacts are carefully documented and well-structured to the extent that reuse and repurposing are facilitated) of research artifacts~\cite{Boisvert2016}.

Artifacts can be software systems, scripts used to run experiments, input datasets, data collected in experiments, or scripts used to analyze results. Artifact evaluation processes help to check their quality via peer review. In some subdisciplines of Computer Science these artifact evaluation processes have been established: Databases (ACM SIGMOD), Software Engineering (ACM SIGSOFT) and Programming Languages (ACM SIGPLAN), see~\cite{Krishnamurthi2015}.  SIGMOD calls the process \textit{reproducibility evaluation} and also offers the `Most Reproducible Paper Award'. The Super Computing Conference Series introduced a reproducibility initiative with an artifact evaluation process in 2016.\footnote{\url{https://sc18.supercomputing.org/submit/sc-reproducibility-initiative/}} 
However, some subdisciplines of Computer Science are still discussing whether they should adopt the artifact evaluation process. Such a subdiscipline is Information Retrieval (ACM SIGIR), which started an initiative to implement the ACM artifact review and badging process~\cite{Ferro2018}. 

Recently, the Empirical Software Engineering journal initiated an open science initiative including an artifact evaluation process \cite{MendezFernandez2019}. Once a manuscript gets \textit{minor revision}, the authors are encouraged to prepare a replication package.
When the manuscript gets accepted, the authors are invited to submit the replication package for evaluation.

Childers and Chrysanthis \cite{ChildersChrysanthis2017} examine how artifact evaluation has incentivized authors, and whether the process is having a measurable impact. They observe a statistical correlation between successfully evaluated artifacts and higher citation counts of the associated papers.
This correlation does not imply a cause-and-effect conclusion, but the hypothesis is that authors who participate in artifact evaluations for whatever reason may have a tendency to be more active and visible in the community.

\section*{The FAIR Principles for Research Software}

The FAIR principles are originally intended to make \textit{data} findable, accessible, interoperable, and reusable~\cite{Wilkinson2016}. However, for open science it is essential to publish research software in addition to research data. Extended to research software, the FAIR principles can be summarized as follows:
\begin{description}
	\item[Findable:] The first step in (re)using data and software is to find it.
	\item[Accessible:] Once the user finds the required data and software, she or he needs to know how to access it, possibly including authentication and authorization if data is involved.
	\item[Interoperable:] The data and software often need to be integrated with other data and software.
	\item[Reusable:] For reusability, metadata, data and software should be well-described such that they can be reused, combined and extended in different settings.
\end{description}
Some communities also use source code itself as data. For example, the Mining Software Repositories community analyzes the rich data available in software repositories to uncover interesting information about software systems and projects~\cite{Hassan2008}.
Data from GitHub, Stackoverflow etc.\ is harvested into repositories such as GHTorrent to be employed in research~\cite{Kotti2019}.
Thus, these principles can also be applied to software, which can be stored and treated as data.

However, at present research software is typically not published and archived using the same practices as FAIR data, with a common vocabulary to describe the artifacts with metadata and in a citable way with a persistent identifier. GitHub is not a platform for archival publishing. Zenodo supports archiving and publishing snapshots from GitHub with persistent DOIs,\footnote{\url{https://guides.github.com/activities/citable-code/}} however, it remains a great challenge to collect, preserve, and share all the software source code.
Research software is the result of creative work that can continue to evolve over time.
In general, software must be continuously maintained to function.

Computer science and software engineering play an important role in the implementation of the FAIR principles, which usually have a focus on helping other disciplines to be FAIR. However, computer science research itself is often also based on software; thus, computer science research software should also consider the FAIR principles.

To analyze the current state of research software publication, we conducted an initial study of research software publication and development behavior, as presented in the following section.

\section*{Relating Research Software to Research Publications}

\begin{figure}[htbp]
	\begin{center}
  \includegraphics[width=\columnwidth]{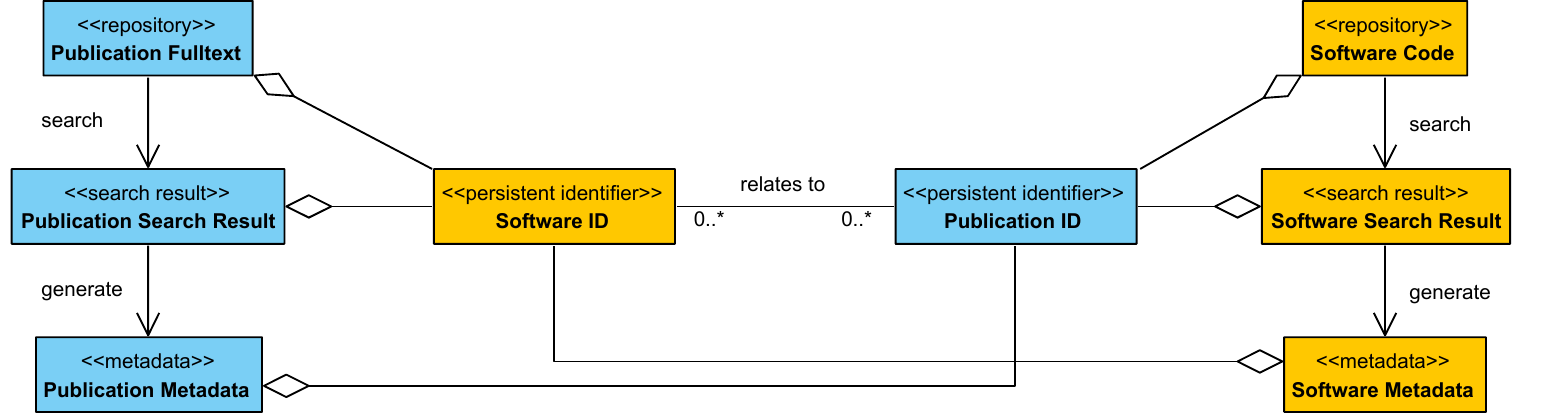}
  \end{center}
	\caption{\label{fig:ResearchSoftwareAnalysis}Conceptual model (as UML class diagram) for our approach to relating research software to research publications. Research publications are searched in their repositories (such as digital libraries). Publication metadata is generated from the search results. The publications and the search results contain the persistent software identifiers as reference to the software repository. The generated metadata contains the publication identifier. A similar process is applied to software code (orange classes). Publication and software identifiers are related in various ways, as will be discussed below.}
\end{figure}

\noindent
To study the relationship between research software and research publications, we conducted an analysis of research software dissemination practices.
For our analysis, research software is identified either by
\begin{itemize}
	\item research publications that cite software repositories or
	\item software repositories that cite research publications.
\end{itemize}
Research software is analyzed in our initial study using a combination of research publication metadata and software repository metadata.
Figure~\ref{fig:ResearchSoftwareAnalysis} illustrates our approach.

\paragraph{Assumptions} We have to make some assumptions for analyzing the relationships between research software and research publications. First, we assume that a research publication refers to some GitHub repository for the related research software. Second, we assume that somewhere in a GitHub repository a publication identifier (DOI) is available. We do \textit{not} assume bi-directional links. We are well aware that these assumptions restrict the coverage of our analysis, but the analysis becomes tractable and repeatable with these assumptions.

\paragraph{Analysis Data Set} Over 5,000 Github software repositories have been identified as \textit{research software} according to the criteria explained previously: either a research publication referenced the software repository, or the software repository referenced
a research publication. This data set is formed from three investigations: (i) 1,204 Github repositories that contain a DOI, (ii) 1,091 Github repositories that are mentioned in publications in the ACM digital library, (iii) 2,872 repositories that are mentioned by e-prints in the arXiv service. In the following section, these will be referred to as the \textit{GitHub}, \textit{ACM} and \textit{arXiv} sets, respectively.

\paragraph{Covered research areas} An first interesting observation is that our three data sets cover quite different research areas:

\begin{itemize}
	\item The GitHub research software set is drawn mainly from the computational sciences, particularly the life sciences (Figure~\ref{fig:TopicsGitHub}). This is determined by resolving the DOI to obtain publication metadata at \url{datacite.org} and classifying the publication venue (e.g.\ journal or conference) in which the paper appeared. The most popular venues were PLOS One, PLOS Computational Biology, Scientific Reports, PNAS, Nature, Nature Communications, Neuroimage, Molecular Biology and Evolution, and Science. 

\begin{figure}[htbp]
\begin{subfigure}{.5\textwidth}
  \centering
  \includegraphics[width=.77\linewidth]{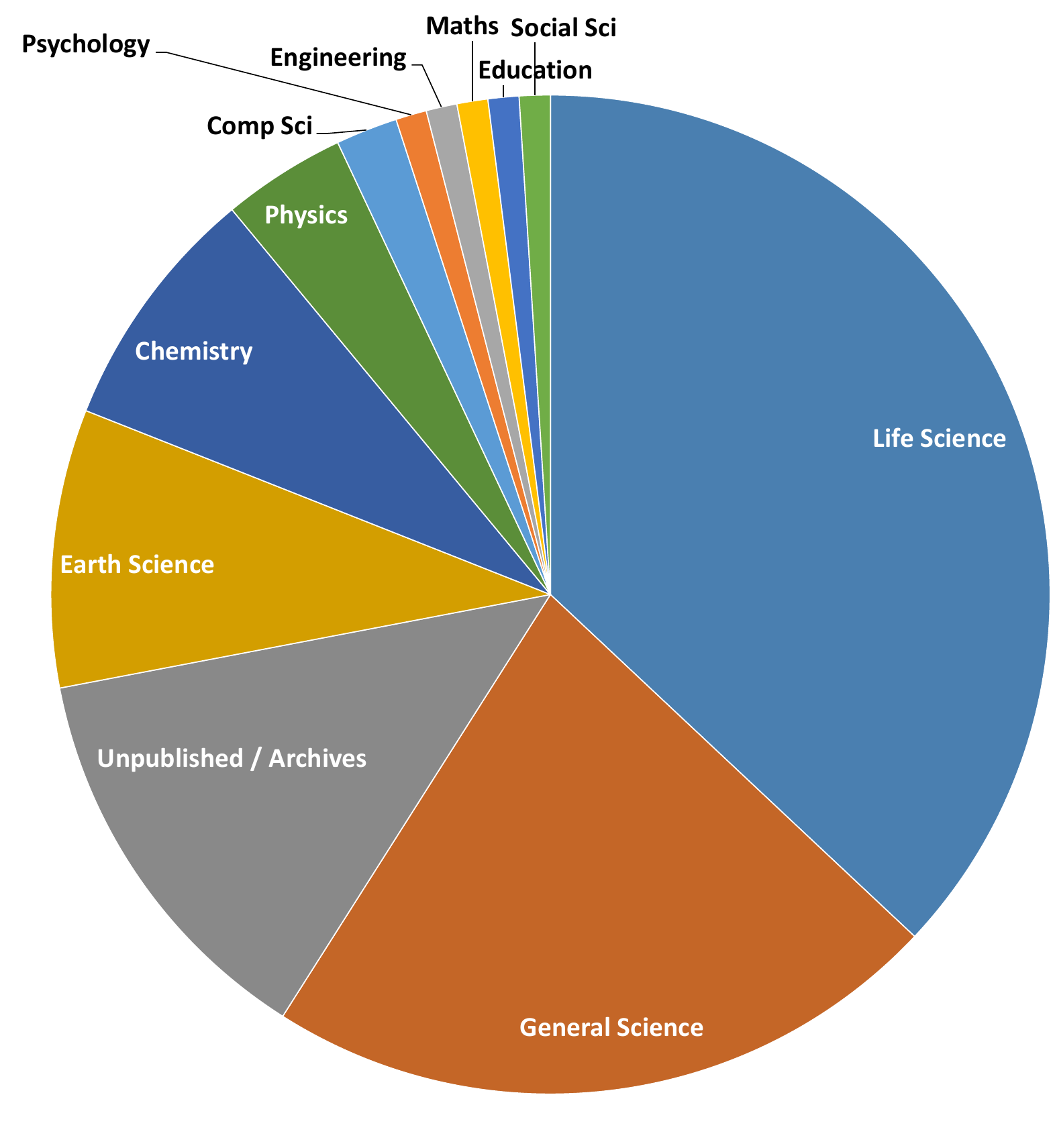}  
  \caption{Research areas of publications cited from\\ Github repositories}
  \label{fig:TopicsGitHub}
\end{subfigure}
\begin{subfigure}{.5\textwidth}
  \centering
  \includegraphics[width=.9\linewidth]{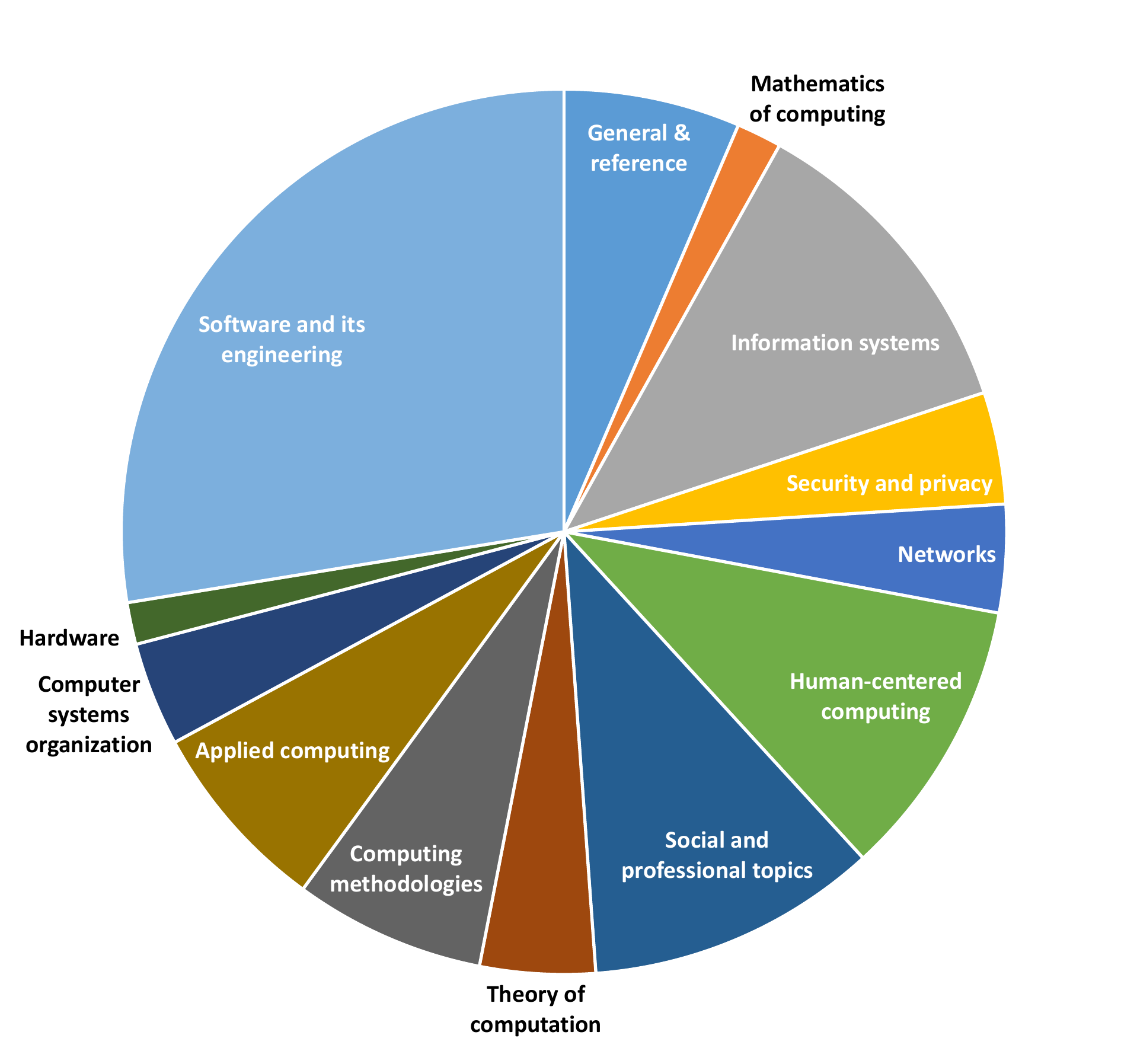}  
  \caption{Research areas of ACM computer science\\ publications citing GitHub repositories}
  \label{fig:TopicsACM}
\end{subfigure}

\vspace{2cm}

\begin{subfigure}{.5\textwidth}
  \centering
  \includegraphics[width=.95\linewidth]{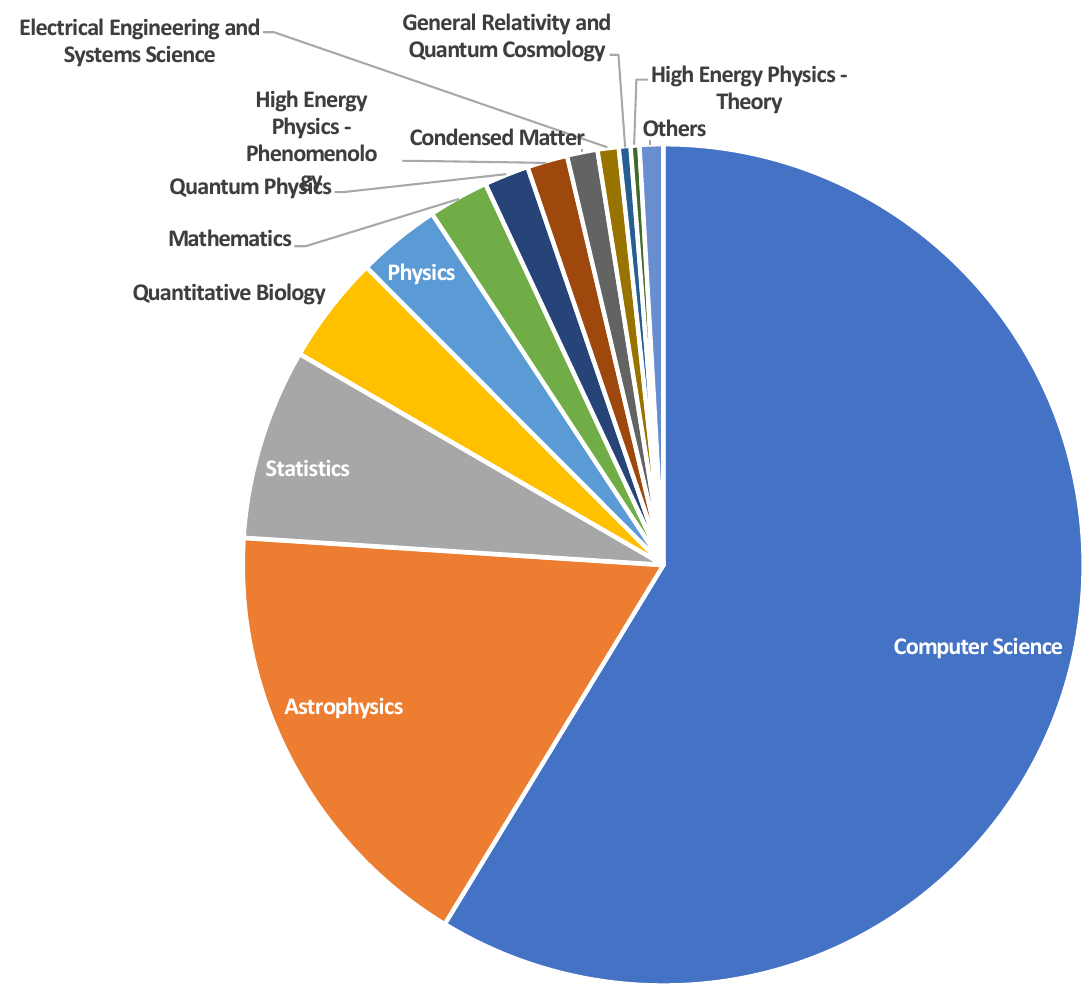}  
  \caption{Research areas of arXiv publications\\ citing GitHub repositories}
  \label{fig:TopicsArXiv}
\end{subfigure}
\begin{subfigure}{.5\textwidth}
  \centering
  \includegraphics[width=.95\linewidth]{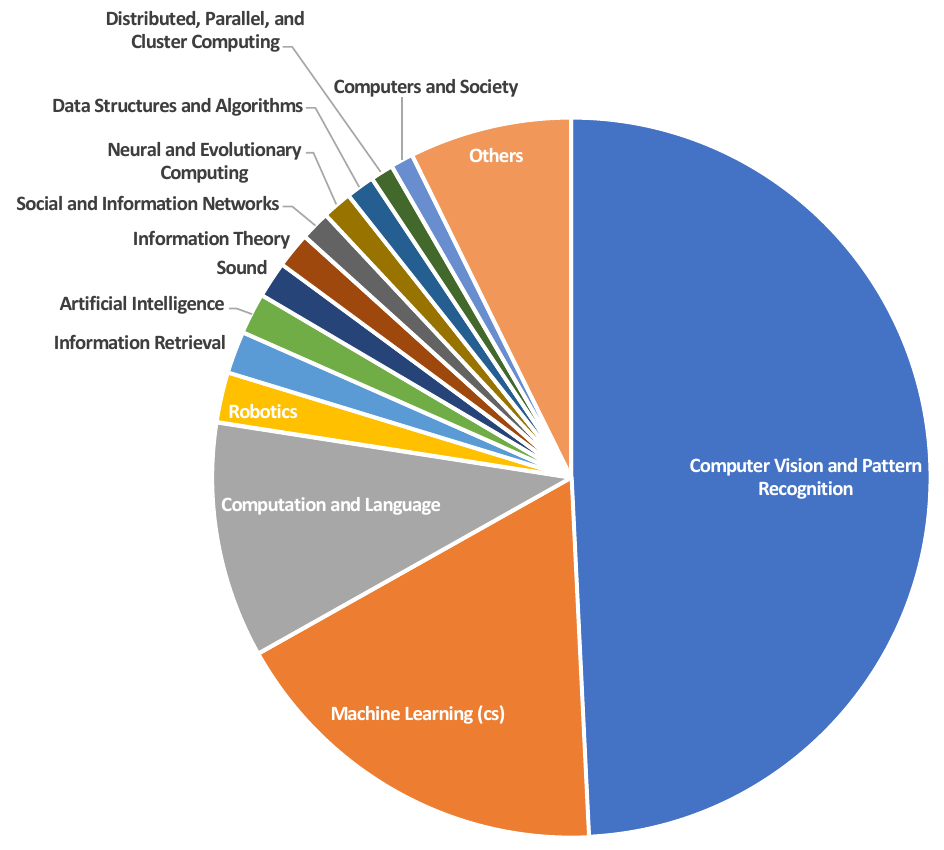}  
  \caption{Computer science publications in arXiv\\ from Figure~\ref{fig:TopicsArXiv} refined into sub-areas}
  \label{fig:TopicsArXivCS}
\end{subfigure}
\caption{\label{fig:Topics}Research areas of publications with related GitHub repositories.}
\end{figure}

	\item The ACM research software set is dominated by software engineering, information systems, social and professional topics and human-centered computing (Figure~\ref{fig:TopicsACM}). This was determined by inspecting the top level of the ACM Computing Classification Scheme (CCS) descriptors which were applied to the publications by the authors.
	\item The arXiv research software set is dominated by computer science topics (Figure~\ref{fig:TopicsArXiv}),\footnote{Remind that we only collect arXiv publications that refer to GitHub repositories.} which is mainly composed of AI topics (computer vision, machine learning, computational linguistics, Figure~\ref{fig:TopicsArXivCS}).  This was determined by inspecting the thematic ``primary category'' to which the e-print was submitted (e.g.\ arXiv categories rcs.AI or hep-th).
Thus, these computer science sub-areas at arXiv seem to emphasize a ``publish and share as early as possible'' attitude, which is encouraged by the review-less publication repository arXiv.
\end{itemize}

\paragraph{Sustainability of research software} 
For this study, we consider research software as sustainable, if it has a greater lifespan and is still live.
We consider a repository as live if some activity occurred during the last year, otherwise it is considered dormant.
The ``lifespan'' of a software repository is the length of time between its first and last commit activity. 
To analyze the sustainability of research software, we divide the software repositories between ``live'' and ``dormant'' repositories. 

\begin{figure}[htbp]
\begin{subfigure}{\textwidth}
  \centering
  \includegraphics[height=.31\textheight]{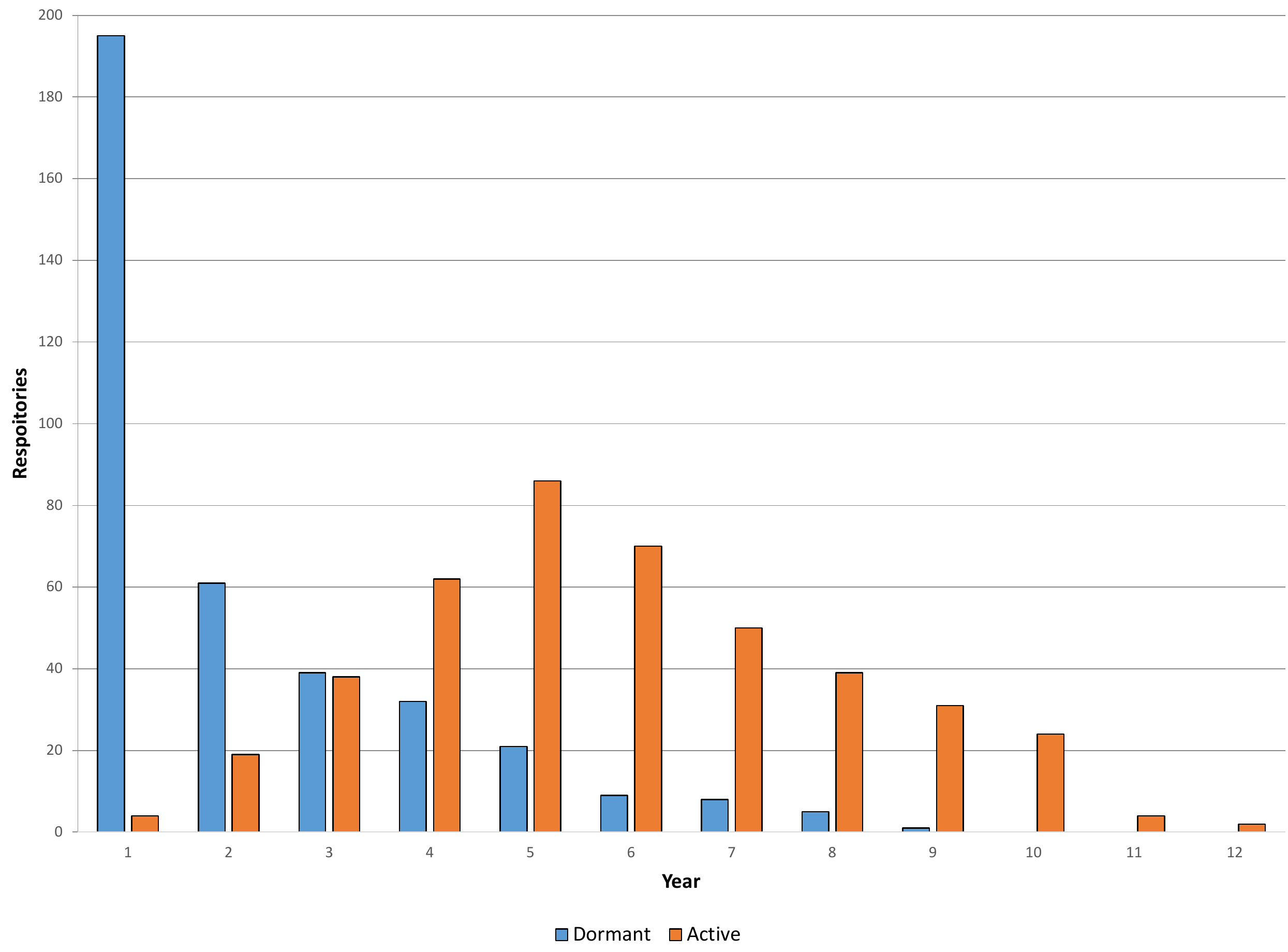}  
  \caption{Lifespan of Github repositories cited in ACM computer science publications}
  \label{fig:LifespanACM}
\end{subfigure}
\begin{subfigure}{\textwidth}
  \centering
  \includegraphics[height=.27\textheight]{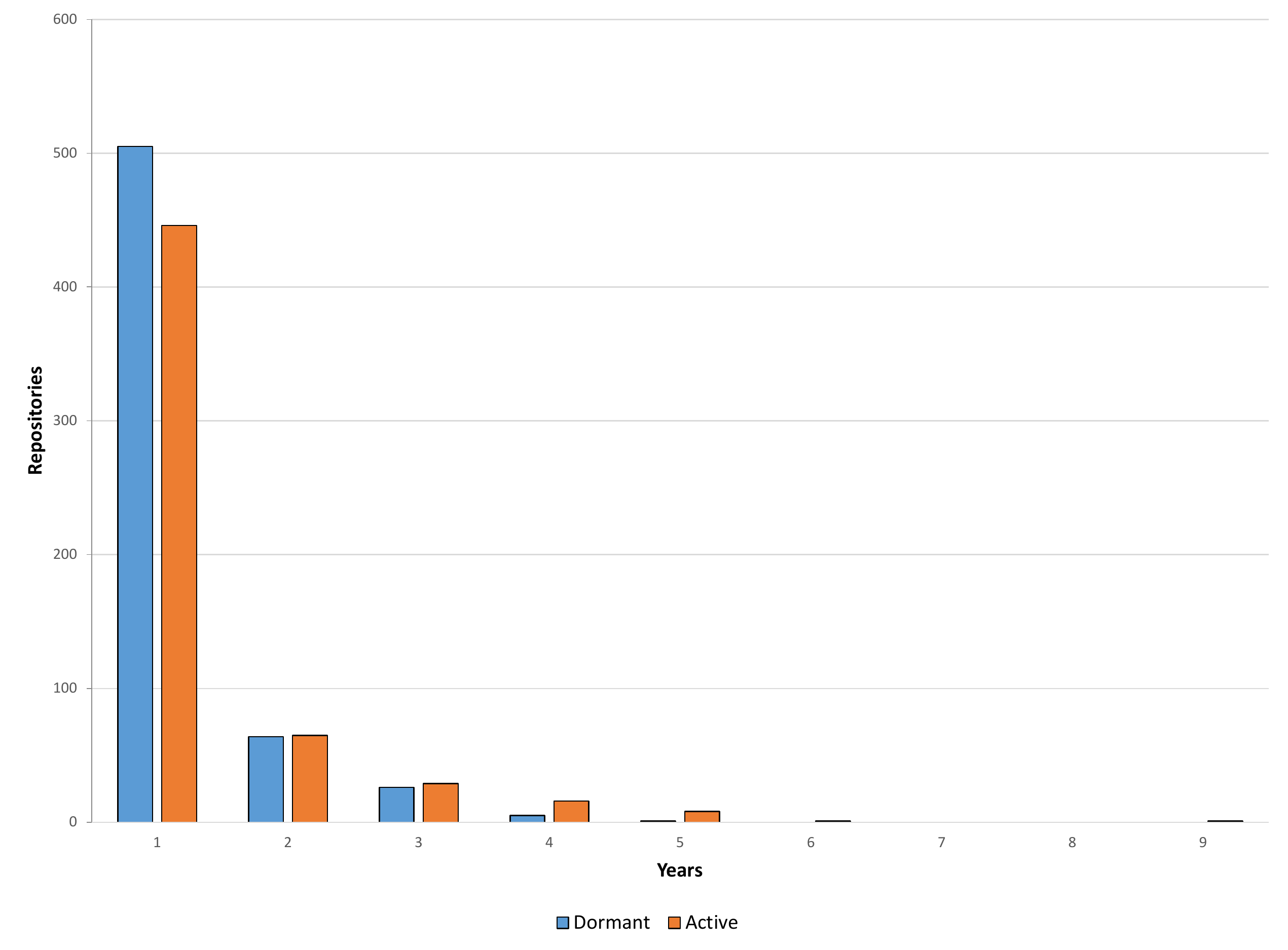}  
  \caption{Lifespan of Github repositories citing publications via DOIs}
  \label{fig:LifespanGitHub}
\end{subfigure}
\begin{subfigure}{\textwidth}
  \centering
  \includegraphics[height=.27\textheight]{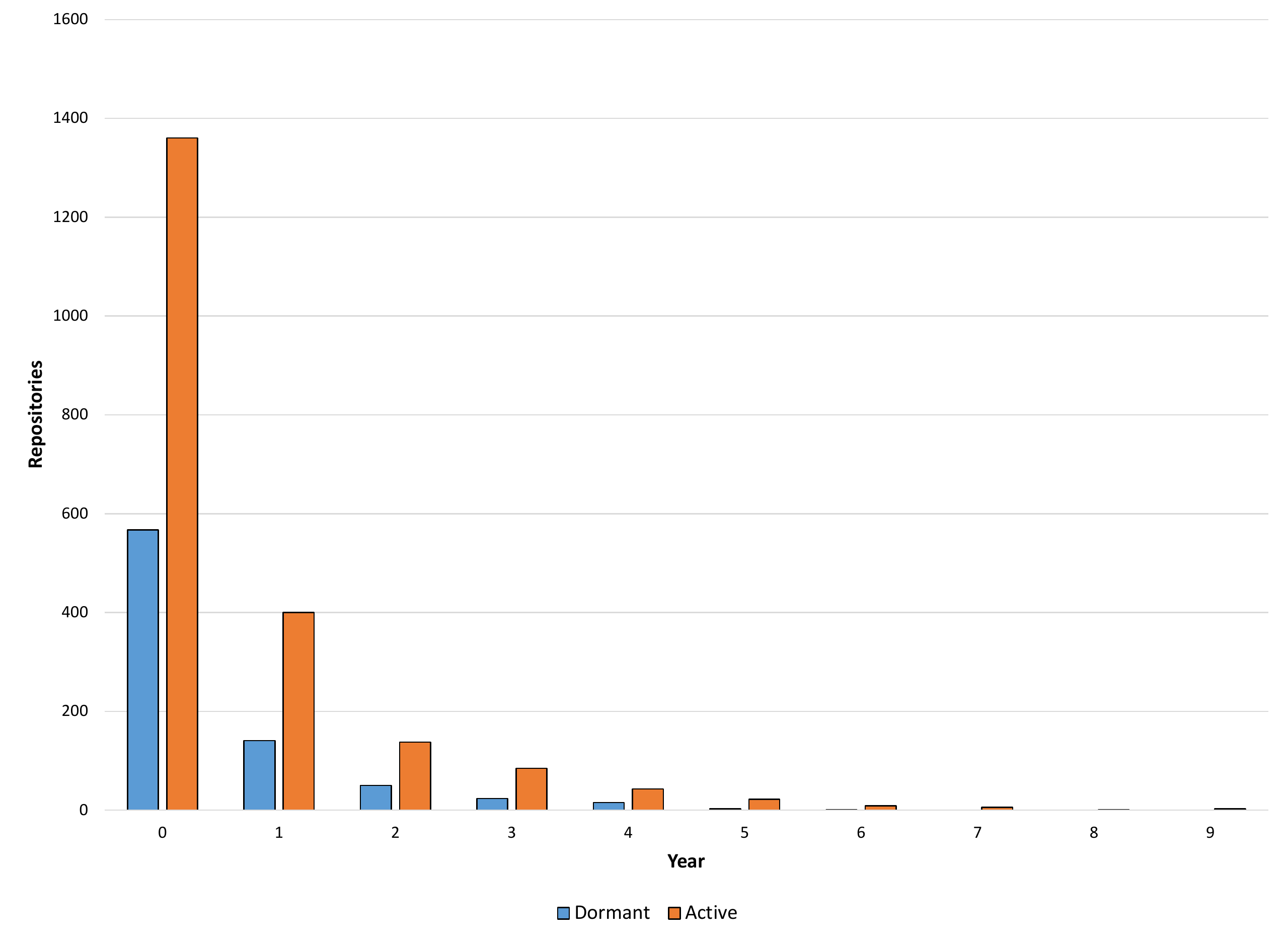}  
  \caption{Lifespan of Github repositories cited in arXiv publications}
  \label{fig:LifespanArXiv}
\end{subfigure}
\caption{\label{fig:Lifespan}Lifespan of software repositories in years.}
\end{figure}

As presented above, publications cited from GitHub repositories mainly belong to computational science, for which we observe an even split between live and dormant software repositories. Publications from the ACM digital library mainly belong to computer science, for their cited software repositories, we also observe an even split between live and dormant software repositories. However, the computer science software repositories lifespan is hugely higher than the computational science software repositories lifespan:

\begin{itemize}
	\item As Figure~\ref{fig:LifespanACM} shows, the computer science software repositories' lifespan is distributed with a median of 5~years.
	
	Our hypothesis is that in computer science research, often commercial open-source software frameworks are employed. These software frameworks are maintained over long times by employees of the associated companies. 
	\item As Figure~\ref{fig:LifespanGitHub} shows, the computational science software repositories' lifespan has a distribution with a median lifespan of 15~days. A third of these repositories are live for less than 1~day.
	
	Our hypothesis is that in computational science research, often the research software is only published when the corresponding paper has been published. The software is then not further maintained at GitHub, but at some private place as before (if it is further maintained at all).
	\item As Figure~\ref{fig:LifespanArXiv} shows, the arXiv repositories are somewhere in between with a median of 8~months lifespan. Furthermore, 75\% of the arXiv repositories are live.
	
	Our hypothesis is that the attitude of publishing as early as possible in parts of the artificial intelligence community also motivates the researchers to develop their research software openly from the start of research projects.
\end{itemize}

\paragraph{Relationships and categories of research software}
In addition to the lifespan, it is interesting to take a closer look at the activity of the repositories; such as the number of commits per time unit. Due to limited space in the present paper, we leave a detailed analysis and discussion of the repositories' activities to future work.
However, by manually inspecting the most active of the ACM repositories, we were able to identify particular kinds of relationship between the research publications and the software repositories, and different kinds of research software.
We observe different categories and relationships between research publications and research software:
\begin{itemize}
	\item Software as an output of research, collaboratively constructed and maintained through an active open source community.
	
	For instance, Caffe is a deep learning framework that has been developed as research software \cite{Jia2014}. It has meanwhile been maintained at GitHub for five years with a large user community and even commercial forks.

	\item Software as an output of research, privately developed but published openly and abandoned after publication.
	
	 For instance, the software for the genetics study by Hough et al.\ \cite{Hough2014} has been published at GitHub\footnote{\url{https://github.com/arvidagren/Cytonuclear}} in 2014 in parallel with the paper. This repository is now five years old with a lifespan of one day (all commits on September 5th, 2014).
	
	\item Software itself as an object of study or analysis.
	
	For instance, Costa et al.\ \cite{Costa2017} studied the performance of the Google Guava core libraries for Java. They did not develop or adapt this software.

	\item Software that then leads to a fork (in GitHub) that is independently developed as a research output and published openly (if successful, it may be fed back into the original project via GitHub pull requests).
	
	For instance, Bosagh Zadeh et al.\ \cite{BosaghZadeh2016} extended the Apache Spark analytics engine as a GitHub fork in their research and managed to merge their software extensions back into the master software repository.

	\item Software used as a tool or framework to do the research.
	
	For instance, O'Donovan et al.\ \cite{O'Donovan2016} used the three.js Javascript 3D library to study 3D modeling approaches.
\end{itemize}
Besides these relationships, software is cited as related work, background, or example. GitHub repositories are also used to publish data and reference lists to collections of software.

\paragraph{Related Work} 
Collberg \&\  Proebsting \cite{Collberg2016} studied the extent to which computer systems researchers share their code and data. Their focus is on re-building the research software for repeatability and on sharing contracts, not FAIR and open publishing. 

Russell et al.\ \cite{Russell2018} conducted a large-scale analysis of bioinformatics software on GitHub looking at relationships between code properties, development activity, and their mentioning in bioinformatics articles. Similar to our observations, they observed that certain scientific topics are associated with more active code development and higher community interest in the repository. Russell et al.\ \cite{Russell2018} focus on bioinformatics research software, while we focus on computer science research software, and the differences in computational science.

The Research Software Directory is a content management system for research software that aims to improve the findability, citability, and reproducibility of the software packages advertised in it, while enabling a qualitative assessment of their impact \cite{ResearchSoftwareDirectory2019}. This related initiative collects research software, but does not analyze the relationship to research publications.

\paragraph{Threats to Validity} As mentioned above, we had to make some assumptions for this initial study. To make our analysis tractable and repeatable, we assume that a research publication refers to some GitHub repository for the related research software or that somewhere in a GitHub repository a publication identifier is available. We are well aware that these assumptions restrict the coverage of our analysis, but even with this limited coverage, we already observed interesting differences in software publication behaviors in different research domains. In our future work, we intend to extend and refine this analysis, for instance to perform a deeper analysis of the repository activities.

Research software is not always cited with a link to the GitHub repository. It could also be published, for instance, in Bitbucket\footnote{\url{https://bitbucket.org}} or GitLab\footnote{\url{https://gitlab.com}} repositories. Alternative citations may refer to papers, manuals or books that introduce the software. Our initial analysis does not cover such additional citation links. To allow for a more comprehensive study of the relationships between research software and research publications, so-called \textit{Research Software Observatories} could provide appropriate citation links and citation graphs, as will be discussed in the following section.

\section*{Observatories for FAIR and Open Research Software}

Based on our experience with analyzing the relationships of research software and research publications, we  propose the deployment of \textit{Research Software Observatories} to better support research software retrieval and analysis.
Discovery and analysis of data resources have been considered in the conceptualization of web observatories \cite{tiropanis_web_2014} and later in data observatories \cite{Tiropanis2019}. A data observatory is a catalogue of data resources and of observations (analytic applications) on those resources. Data observatories envisage decentralized, interoperable catalogues of data resources, hosted by different organizations.\footnote{Reference implementations have already emerged; e.g.\ \url{https://github.com/webobservatory/}} Thus, data observatories are distributed, federated collections of datasets and tools for analyzing data, each with their own user community. Decentralization in this sense can provide for agile architectures in ways that centralized, one-size-fits-all solutions cannot.

In the context of open science and research software, research software observatories can be considered in three different ways. First, in terms of describing a research software observatory for FAIR and open research software, that will allow scientists to share software and observations on the status of this software, such as those described in this publication and illustrated in Figures~\ref{fig:Topics} and~\ref{fig:Lifespan}. A research software observatory could support open science research and encourage best practice among research communities. Second, one could consider the research software used for processing scientific data and producing observations (analytics) in ways that respect the FAIR and open principles. Third, the opportunities and challenges of cataloging research software with appropriate citation links in observatories can be explored. Research software observatories need to support metadata for research software classification and citation to further empower researchers to find, access and reuse relevant and interoperable research software.

\section*{Recommendations to make Computer Science Research Software FAIR and Open}

Publishing research software in an archival repository is currently not common in all areas of computer science. Our initial study revealed highly varying publication behavior in different scientific disciplines. Research software is usually managed in GitHub or similar repositories, where it can be maintained and re-used, but not published for scientific reward and proper citation.
An approach to addressing these issues is by enabling and standardizing citation of software. Software citation brings the effort of software development into the current model of academic credit, while simultaneously enhancing the scholarly record by linking together software with publications, datasets, methods, and other research objects.
Therefore, our recommendations along to the FAIR principles are the following:
\begin{description}
	\item[For findability,] challenges to be addressed for FAIR publication of research software are methods for software citation and software retrieval. To support findability, computer science sub-disciplines may adopt approaches that are currently under exploration for research software in general. However, appropriate software metadata remains a great challenge. 
	
Authors sometimes want their users to cite something other than the piece of software directly. Examples include citing a paper that introduces the software, a published software manual or book, a `software paper' (such as JOSS) created specifically as a citation target, or a benchmarking paper.

However, there exists guidelines for software citation and identification~\cite{FORCE11}, and already some metadata standards for software citation exist~\cite{Katz2018Springer}:
\begin{itemize}
	\item The Citation File Format (CFF) is a human- and machine-readable file format in YAML which provides citation metadata for software~\cite{Druskat2018}.
	\item A CodeMeta instance file describes the metadata associated with a software object using JSON's linked data (JSON-LD) notation~\cite{Boettiger2017}.
\end{itemize}
The \url{CiteAs.org} online service links between software repositories and their requested citations, exploiting the above standards. What is missing, are search engines that exploit this metadata and, more importantly, widespread annotation of research software with citation information.

\newpage
	\item[For accessibility,] software artifacts should be published with preservation in mind. GitHub, for example, does not directly support the preservation of software ``snapshots'' which were used to achieve some research results. This may, for instance be achieved via taking a snapshot from GitHub to be archived on Zenodo.org:\footnote{\url{https://guides.github.com/activities/citable-code/}} 
\begin{itemize}
	\item GitHub serves for use, reuse, and active involvement of researchers.
	\item Zenodo serves for archival and reproducibility of published research results.
\end{itemize}
An open question is whether computer science research needs its own discipline-specific data repository and whether the combination of GitHub and Zenodo is sufficient.
The Software Heritage archive could be another option for software preservation~\cite{dicosmo2017}.

	\item[For interoperability,] research software engineers should adhere to established software and data standards allowing for interoperable software components \cite{Standards2000}. Proper interface definitions in modular software architectures are essential for interoperable research software components.

Artifact evaluation processes may support interoperability, if the reviewers take this concern into account.

	\item[For reusability,] artifact evaluation processes review replicability and reproducibility and, if successful, reusability of research software. This way, the reusability of research software may be improved significantly.

Software virtualization techniques such as Docker containers and online services help to support portability, and thus reusability across platforms. It may be useful to distinguish between Software-as-Code (e.g., via GitHub) and Software-as-a-Service (e.g., via some online cloud service on which the software is executed, such as BinderHub). 
	
	From a software engineering point of view, modular software architectures allow for reusing parts of research software systems 
	\cite{SA2018}. So far, many research software systems are not structured in a modular architecture, what should be improved in the future. Domain-specific languages may also help with the comprehensibility of research software \cite{ESE2017}.
	
	It is vital for reusability to follow good engineering practices to ensure that the software can be built on by others \cite{Collberg2016}. Adequate documentation is important, but so are engineering practices such as providing testing frameworks and test data for continuous integration to ensure that future adaptations can be tested to ensure that they work correctly.
\end{description}

\section*{Summary}

Compared to research data, research software should be both archived for reproducibility and actively maintained for reusability. The combination of Zenodo (for archival and reproducibility) and GitHub (for maintenance and reuse) may be used to achieve this.
Furthermore, research software should be open source software. Established open source software licenses~\cite{Ballhausen2019} provide adequate licensing options such that there is no need to keep research software closed. In the vast majority of cases some existing license will be appropriate. For research data this is different. Research data may, for instance, be subject to privacy regulations. Thus, the FAIR \textit{data} principles do not require openness, but accessibility that might include authentication and authorization. However, for research \textit{software}, openness is to be expected \cite{Katz2018}. Only in exceptional cases and for very good reasons should research software be closed.

Reproducibility and reusability are essential for good scientific practice.
Future work should address the definition and establishment of appropriate metadata for citing both software code and software services. Such metadata could make research software also better searchable and discoverable. 
Research software observatories may provide such services for software retrieval and analysis.

Modularity is essential for maintainability, scalability and agility, but also for reusability. 
We suggest to further establish the concept of artifact evaluation to ensure the quality of published artifacts for better reusability.

Proper research software engineering enables reproducibility and reusability of research software in computational science.
However, software engineering should also help software engineering and computer science research itself to support replicability and reproducibility of research software that is used in computer science experiments. 
This way, we may achieve FAIR and open computer science and software engineering research.

\bibliographystyle{plainurl}

\bibliography{references}

\newpage

\noindent
\textbf{Wilhelm Hasselbring} (hasselbring@email.uni-kiel.de) is a full professor of software engineering in the Department of Computer Science at Kiel University, Germany.

\bigskip

\noindent
\textbf{Leslie Carr} (lac@ecs.soton.ac.uk) is a full professor of web science in the Department of Electronics and Computer Science at the University of Southampton, UK.

\bigskip

\noindent
\textbf{Simon Hettrick} (sjh@ecs.soton.ac.uk) is deputy director of UK's Software Sustainability Institute and a full professor in the Department of Electronics and Computer Science at the University of Southampton, UK.

\bigskip

\noindent
\textbf{Heather Packer} (hp3@ecs.soton.ac.uk) is a New Frontier fellow in the Department of Electronics and Computer Science at the University of Southampton, UK.

\bigskip

\noindent
\textbf{Thanassis Tiropanis} (t.tiropanis@southampton.ac.uk) is an associate professor in the Department of Electronics and Computer Science at the University of Southampton, UK.

\end{document}